\documentclass[aps, a4paper,superscriptaddress, nofootinbib, showpacs, twocolumn, 10pt]{revtex4}
\usepackage{amsmath,amssymb,bm,natbib}
\usepackage{epsfig}
\usepackage{graphicx}
\usepackage{slashed}
\usepackage{gensymb}

\newcommand{\beq}{\begin{equation}}
\newcommand{\eeq}{\end{equation}}
\newcommand{\bea}{\begin{eqnarray}}
\newcommand{\eea}{\end{eqnarray}}
\newcommand{\ba}{\begin{align}}
\newcommand{\ea}{\end{align}}
\newcommand{\bfig}{\begin{figure}}
\newcommand{\efig}{\end{figure}}

\newcommand{\D}{\displaystyle}

\newcommand{\gev}{\, \text{GeV}}

\newcommand{\LO}{\, \text{LO}}

\newcommand{\tin}{t_{\rm in}}
\newcommand{\la}{\langle}
\newcommand{\ra}{\rangle}

\newcommand{\omnes}{{\cal{O}}}

\newcommand{\fmsq}{\, \text{fm}^2}

\begin{document}
\begin{flushright}
SI-HEP-2013-16, DO-TH 13/35, QFET-2013-13\\
%DO-TH 13/35\\
%QFET-2013-13
\end{flushright}
\phantom{}
\vspace*{-17mm}

\title{Two-pion low-energy contribution to the muon $g-2$ 
with improved precision from analyticity and unitarity}

\author{B.Ananthanarayan}
\affiliation{Centre for High Energy Physics,
Indian Institute of Science, Bangalore 560 012, India}
\author{Irinel Caprini}
\affiliation{Horia Hulubei National Institute for Physics and Nuclear Engineering,
P.O.B. MG-6, 077125 Magurele, Romania}
\author{ Diganta Das}
\affiliation{Institut f\"ur Physik, Technische Universit\"at Dortmund, D-44221
Dortmund, Germany}
\author{I. Sentitemsu Imsong}
\affiliation{Theoretische Physik 1, Naturwissenschaftlich-Technische Fakult\"{a}t, 
 Universit\"{a}t Siegen, D-57068 Siegen, Germany}

\begin{abstract}
The two-pion contribution from low energies to the
muon magnetic moment anomaly, although small, has a large relative
uncertainty since in this region the experimental data on the cross sections
are neither sufficient nor precise enough.  It is therefore of interest to see 
whether the precision can be improved by means of additional theoretical 
information on the  pion electromagnetic form factor, 
which controls the leading-order contribution. 
In the present paper we address this problem by exploiting 
analyticity and unitarity of the form factor in a parametrization-free 
approach that uses  the phase  in the elastic region, known with  high
precision from the Fermi--Watson theorem and Roy equations for $\pi\pi$ 
elastic scattering as input. The formalism also includes experimental 
measurements on the  modulus in the region 0.65-0.70 GeV, 
taken from the most recent  $e^+e^-\to \pi^+\pi^-$ experiments, 
and recent measurements of the form factor  on the  spacelike axis. 
By combining the results obtained with inputs from CMD2, SND, BABAR and KLOE, 
we make the predictions $a_\mu^{\pi\pi, \LO}\,[2 m_\pi,\, 0.30 \gev]=(0.553 \pm 0.004) \times 10^{-10}$ and $a_\mu^{\pi\pi, \LO}\,[0.30 \gev,\, 0.63 \gev]=(133.083  \pm 0.837)\times 10^{-10}$.
These are consistent with the other recent determinations, 
and have slightly smaller errors.  
\end{abstract}

\maketitle
\section{Introduction} \label{sec:Intro}

The muon anomalous magnetic moment $a_\mu = g_\mu/2 -1$ is one
of the most precisely measured observables in particle physics. 
It can be predicted also by theory with a high accuracy, serving as
a monitor 
for precise tests of the Standard Model (SM) \cite{PDG}.  
The Brookhaven muon $g-2$ experiment \cite{Gm2} revealed a persisting
discrepancy between theory and experiment at the 3 to 4$\sigma$ level.  
The present experimental precision is $\delta_\mu^{\rm exp}\sim  
63 \times 10^{-11}$, while most advanced
theoretical predictions claim an of accuracy 
$\delta_\mu^{\rm th}\sim  49 \times 10^{-11}$ (cf. the recent reviews in Refs.
\cite{Miller:2007kk,Miller:2012opa, PaMa, JeNy, Jegerlehner:2007xe,Gnendiger:2013pva}). 
 
The next generation experiment planned at Fermilab,  
with the goal of reaching a precision
of  $\delta_\mu^{\rm exp}\sim 16 \times 10^{-11}$, strongly demands 
improved theoretical  predictions. At present, the biggest 
theoretical uncertainties are due to the nonperturbative hadronic
contributions, which cannot be calculated from first principles   
\cite{Davier:2009,Davier:2010}. In particular, the leading-order 
(LO) hadronic contribution to vacuum polarization is responsible 
for an uncertainty $\delta_\mu^{\rm LOVP}\sim 41 \times 10^{-11}$, 
of which  about a half  comes from the two-pion contribution 
\cite{Davier:2009}. High statistics cross section measurements 
for the $e^+e^-\to\pi^+\pi^-$ reaction were performed recently by 
CMD2 \cite{CMD2,CMD2:2}, SND \cite{SND}, BABAR \cite{BABAR,BABAR1} 
and KLOE \cite{KLOE1,KLOE2,KLOE3} experiments.  However, there are 
some discrepancies between   the recent experiments, in particular 
BABAR and KLOE  \cite{Bena,DaMa}. Moreover,
the experimental data are not very accurate or are missing at low energies, 
so that the low-energy contribution to the muon  $g-2$, although 
small, has a large uncertainty \cite{Davier:2009}. Improving this 
contribution by exploiting low-energy effective theories for hadrons is therefore highly desirable.

At leading order the contribution of interest can be evaluated in terms of the pion
electromagnetic form factor. An accurate knowledge of this quantity plays a crucial role for the SM calculation of the muon $g-2$. Since the modulus of the form factor at low energies is poorly known, one can instead use the phase, which is related to the modulus by analyticity.
By the Fermi--Watson theorem, in the isospin limit the phase of the pion electromagnetic form factor on the unitarity cut below the first inelastic threshold is equal to the  $P$-wave phase shift of pion-pion scattering. This phase shift was calculated recently with high precision from chiral perturbation theory (ChPT) and Roy equations \cite{ACGL,Caprini:2011ky,KPY,GarciaMartin:2011cn}.
As shown  in Ref.\cite{IC}, it is possible to implement the phase in the elastic region into what has come to be known as the Meiman interpolation problem \cite{Meiman,Duren}, which amounts to finding bounds on an analytic function and its derivatives at points inside the holomorphy domain (for a recent review of these techniques, see \cite{Abbas:2010EPJA}). 

In a series of recent papers 
\cite{Ananthanarayan:2011,Ananthanarayan:2012,Ananthanarayan:2012tt,Ananthanarayan:2013dpa}
we have applied the formalism for improving the knowledge on the pion form factor. Bounds on the form factor in the spacelike region were derived in Ref. \cite{Ananthanarayan:2012} in order to test the onset of the asymptotic behavior  predicted by perturbative QCD. We have also derived stringent constraints  on  the shape parameters (at the origin) \cite{Ananthanarayan:2011, Ananthanarayan:2013dpa} and  on the modulus in the near threshold part of the unitarity cut \cite{Ananthanarayan:2012tt},  which superseded the experimental data in accuracy.  We emphasize that the method does not rely on specific models of the form factor, being a parametrization-free approach. The price to be paid is that we obtain bounds rather than definite values for the quantities of interest. However, with the increased precision of input available now,  the bounds are quite stringent, competing with specific models and experiment  within errors. 

In the present work we explore the consequences of the formalism for the muon $g-2$. We have already noted that the bounds on the modulus calculated in Ref. \cite{Ananthanarayan:2012tt} lead to a more precise description of the modulus at low energies than the experimental data. We now further improve these bounds by using some experimental values on the modulus measured at higher energies as input, where the precision is better and the data from various experiments are more consistent among themselves. The aim is to establish whether the calculated bounds on $|F(t)|$ are able to improve the accuracy of the hadronic part of muon $g-2$.

The scheme of this paper is as follows. In Sec. \ref{sec:aim}  
we briefly review  the basic formulas that set the stage for our work. In Sec. \ref{sec:problem}  we formulate the extremal problem that plays the crucial role in our formalism, and in Sec. \ref{sec:sol} we give the solution of this problem.  In Sec. \ref{sec:inputs} we discuss the 
input that goes into our analysis and explain how the various uncertainties are taken into account.  In Sec. \ref{sec:results}, we present our results for the low-energy
pionic contribution to the muon $g-2$  
and compare them with previous determinations. 
Section \ref{sec:conclusion} contains our conclusions.

%%%%%%%%%%%%%%%%%%%%%%%%%%%%%%%%%%%%%%%%%%%%%%%%%%%%%%%%%%%%%%%%%%%%%%%%%%%%%%%%%%%%%%%%%%%%
\section{Basic formulas \label{sec:aim}}
%%%%%%%%%%%%%%%%%%%%%%%%%%%%%%%%%%%%%%%%%%%%%%%%%%%%%%%%%%%%%%%%%%%%%%%%%%%%%%%%%%%%%%%%%%%%
As in the recent experimental works \cite{CMD2, CMD2:2,  SND, BABAR,  BABAR1,  KLOE1,  KLOE2, KLOE3}, we consider the LO two-pion contribution to $a_\mu$, which does not contain the
vacuum polarization effects but includes one-photon final-state radiation (FSR). It is expressed in terms of the pion electromagnetic form factor $F(t)$ as
\begin{equation} \label{eq:amu}
a_\mu^{\pi\pi, \LO} = \frac{\alpha^2 m_\mu^2}{12 \pi^2}\int_{t_+}^\infty \frac{dt}{ t} \, K(t)\, \beta^3_\pi(t) \,   
|F(t)|^2 \left(1+\frac{\alpha}{\pi}\,\eta_\pi(t)\right),
\end{equation}
where $t_+=4m_\pi^2$, $\beta_\pi(t)=(1-t_+/t)^{1/2}$ and
\begin{equation}
K(t) = \int_0^1 du(1-u)u^2(t-u+m_\mu^2u^2 )^{-1}.
\end{equation}
The last factor in (\ref{eq:amu})  accounts
for the  FSR, calculated in scalar QED \cite{FSR1,FSR2}.

 We emphasize that the form factor $F(t)$ is defined by
\beq\label{defe}  \langle \pi^+(p')\vert J_\mu^{\rm elm} \vert \pi^+(p)\rangle= (p+p')_\mu F(t), ~ t=(p-p')^2,
\eeq
such as to satisfy the Fermi--Watson theorem.  Since the experimental collaborations (CMD2, SND, BABAR, and KLOE) include the vacuum polarization into the definition of the pion form factor, to obtain $|F(t)|$ from experiment   we remove the vacuum polarization  from the values of the modulus quoted in Refs. \cite{CMD2, CMD2:2,  SND, BABAR,  BABAR1,  KLOE1,  KLOE2, KLOE3}. Equivalently, we extract  $|F(t)|$  directly from the measured cross section by
\begin{equation}
|F(t)|^2=\frac{3t}{\alpha^2 \pi \beta_\pi(t)^3} \frac{\sigma^0_{\pi\pi(\gamma)}(t)}{1+\frac{\alpha}{\pi}\,\eta_\pi(t)},
\end{equation}
where $\sigma^0_{\pi\pi(\gamma)}$ is the undressed cross section of $e^+e^-\to \pi^+\pi^-(\gamma)$  quoted in Refs. \cite{CMD2, CMD2:2,  SND, BABAR,  BABAR1,  KLOE1,  KLOE2, KLOE3}.

We are interested in finding the low-energy part of the integral (\ref{eq:amu}). For comparison with previous works \cite{Davier:2009} we shall evaluate  in particular the contributions $a_\mu^{\pi\pi, \LO}[2 m_\pi, 0.3\gev]$ and  $a_\mu^{\pi\pi, \LO}[0.3\gev, 0.63\gev]$. As we have mentioned, at these energies the experimental data on $|F(t)|$ are scarce and have rather large errors. Therefore, we shall replace them with upper and lower bounds on $|F(t)|$ calculated from the extremal problem to be formulated in the next section.

%%%%%%%%%%%%%%%%%%%%%%%%%%%%%%%%%%%%%%%%%%%%%%%%%%%%%%%%%%%%%%%%%%%%%%%%%%%%%%%
\section{Extremal problem \label{sec:problem}}
%%%%%%%%%%%%%%%%%%%%%%%%%%%%%%%%%%%%%%%%%%%%%%%%%%%%%%%%%%%%%%%%%%%%%%%%%%%%%%%
We consider the following conditions on $F(t)$:
\vskip0.2cm
 1.  The Fermi--Watson theorem:
\beq\label{eq:watson}
{\rm Arg} [F(t+i\epsilon)]=\delta_1^1(t),  \quad\quad t_+ \le t \le \tin,
\eeq
where $\delta_1^1(t)$ is the phase shift of the $P$-wave of $\pi\pi$ elastic scattering and 
$\tin$ is the first inelastic threshold in the unitarity sum.

 2.  An integral condition on the modulus squared above the inelastic threshold, written in the form
\beq\label{eq:L2}
 \D\frac{1}{\pi} \int_{\tin}^{\infty} dt \rho(t) |F(t)|^2 \leq  I,
 \eeq
where $\rho(t)$ is a suitable positive-definite weight, for which the integral converges and an accurate evaluation of $I$ is possible. 

 3. The known  first two Taylor coefficients at $t=0$:
\beq\label{eq:taylor}
	F(0) = 1, \quad  \quad \left[\frac{dF(t)}{dt}\right]_{t=0} =\D\frac{1}{6} \la r^2_\pi \ra.
\eeq

 4. The value at one  spacelike energy: 
\beq\label{eq:val}
F(t_s)= F_s \pm \epsilon_s, \qquad t_s<0.
\eeq

  5. The value of the modulus at one energy in the elastic region of the timelike axis: 
\beq\label{eq:mod}
|F(t_t)|= F_t \pm \epsilon_t, \qquad t_+< t_t <\tin.
\eeq

\vskip0.2cm
We now formulate the following problem: {\em find optimal upper and lower bounds on $|F(t)|$ on the elastic unitarity cut, $t_+<t<\tin$  for $F(t)\in {\cal C}$, where  ${\cal C}$ is the class of functions real analytic in the $t$ plane cut along the real axis for $t\ge t_+$,  which satisfy the conditions 1--5 given
above.}
%%%%%%%%%%%%%%%%%%%%%%%%%%%%%%%%%%%%%%%%%%%%%%%%%%%%%%%%%%%%%%%%%%%%%%%%%%%%%%%
\section{Solution}\label{sec:sol}
%%%%%%%%%%%%%%%%%%%%%%%%%%%%%%%%%%%%%%%%%%%%%%%%%%%%%%%%%%%%%%%%%%%%%%%%%%%%%%%
For solving the extremal problem stated above, we use a mathematical method presented in \cite{IC,Abbas:2010EPJA}.
 We first define the Omn\`{e}s function
\beq	\label{eq:omnes}
 \omnes(t) = \exp \left(\D\frac {t} {\pi} \int^{\infty}_{t_+} dt' 
\D\frac{\delta (t^\prime)} {t^\prime (t^\prime -t)}\right),
\eeq
where $\delta(t)=\delta_1^1(t)$   for 
$t\le \tin$, and is an arbitrary function, sufficiently  smooth ({\em i.e.,}
Lipschitz continuous) for $t>\tin$. As discussed in detail in Ref.  \cite{Abbas:2010EPJA}, the results do not depend on the 
choice of the function  $\delta(t)$ for $t>\tin$.

 We remark that the function $h(t)$ defined by
\beq\label{eq:h}
F(t)=\omnes(t) h(t)
\eeq
is analytic in the $t$-plane cut only for $t>\tin$. 
In terms of $h(t)$ the equality (\ref{eq:L2}) writes as
\beq\label{eq:hL2}
\D\frac{1}{\pi} \int_{\tin}^{\infty} dt\, 
\rho(t) |\omnes(t)|^2 |h(t)|^2 \leq  I.
\eeq

This relation is written in a canonical form if we perform the conformal transformation
\beq\label{eq:ztin}
\tilde z(t) = \frac{\sqrt{\tin} - \sqrt {\tin -t}} {\sqrt{\tin} + \sqrt {\tin -t}}\,,
\eeq
which maps the complex $t$-plane cut for $t>\tin$  onto the unit disk $|z|<1$ in the $z$ plane defined 
by $z\equiv\tilde z(t)$, such that the origin $t=0$ of the $t$ plane is mapped onto the origin $z=0$ of the $z$ plane, the point $t=\tin$ becomes $z=1$ and the upper/lower edges of the cut along $t>\tin$ become the upper/lower halves of the unit circle $\zeta=\exp(i\theta)$. We further define a function $g(z)$ by
\beq	\label{eq:gF11}
 g(z) = w(z)\, \omega(z) \,h(\tilde t(z)).
\eeq 
In this relation  $\tilde t(z)$ is the inverse of $z = \tilde z(t)$, for $\tilde z(t)$  defined in
Eq.(\ref{eq:ztin}), and $w(z)$ and $\omega(z)$ are outer functions, {\it i.e.} functions analytic and without zeros in 
the unit disk $|z|<1$, defined in terms of their modulus on the boundary $|z|=1$, related to  
$\sqrt{\rho(t)\, |{\rm d}t/ {\rm d} \tilde z(t)|}$ 
and  $|\omnes(t)|$,  respectively \cite{IC, Abbas:2010EPJA}.
In particular, choosing in Eq.(\ref{eq:L2})  weight functions $\rho(t)$ of the form 
\beq \label{eq:rhogeneric0}
\rho(t) = \frac{t^b}{(t+Q^2)^c}, \quad \quad  Q^2\ge0,~  b\leq c \leq b+2,
\eeq the first outer function, $w(z)$,  
can be written in an analytic closed form in the $z$ variable as \cite{Abbas:2010EPJA}
\beq\label{eq:outerfinal0}
w(z)= (2\sqrt{t_{\rm in}})^{1+b-c}\frac{(1-z)^{1/2}} {(1+z)^{3/2-c+b}}\frac{(1+\tilde z(-Q^2))^c}{(1-z \tilde z(-Q^2))^c}.
\eeq 
For the second outer function, denoted as $\omega(z)$,  we use an integral representation in terms 
of its modulus on the cut $t>\tin$, which can be written as \cite{IC, Abbas:2010EPJA}
\beq\label{eq:omega}
 \omega(z) =  \exp \left(\D\frac {\sqrt {\tin - \tilde t(z)}} {\pi} \int^{\infty}_{\tin}  \D\frac {\ln |\omnes(t^\prime)|\, {\rm d}t^\prime}
 {\sqrt {t^\prime - \tin} (t^\prime -\tilde t(z))} \right).
\eeq 

 Since  the function $h(\tilde t(z))$ defined in Eq.(\ref{eq:h})  is analytic in 
$|z|<1$,  it follows that the function $g(z)$ itself is analytic in   $|z|<1$. Moreover,   the relation  (\ref{eq:hL2}) is written in terms of $g(z)$ as
\beq\label{eq:gI1}
\frac{1}{2 \pi} \int^{2\pi}_{0} {\rm d} \theta |g(\zeta)|^2 \leq I, \qquad \zeta=e^{i\theta}.
\eeq
As proven in the so-called analytic interpolation theory \cite{Meiman,Duren}, the  $L^2$-norm condition (\ref{eq:gI1}) leads to rigorous correlations among the values of the analytic function  $g(z)$ and its derivatives at points inside the holomorphy domain, $|z|<1$. In particular, one can show (for a proof and earlier references, see Ref. \cite{Abbas:2010EPJA}) that Eq.(\ref{eq:gI1}) implies the positivity
\beq\label{eq:posit}
{\cal D}\ge 0
\eeq
 of the determinant ${\cal D}$ defined as
\beq\label{eq:det}
{\cal D}=\left|
\begin{array}{c c c c c c}
\bar{I} & \bar{\xi}_{1} & \bar{\xi}_{2} & \ldots & \bar{\xi}_{N}\\	
	\bar{\xi}_{1} & \D \frac{z^{2K}_{1}}{1-z^{2}_1} & \D
\frac{(z_1z_2)^K}{1-z_1z_2} & 	\ldots & \D \frac{(z_1z_N)^K}{1-z_1z_N} \\
	\bar{\xi}_{2} & \D \frac{(z_1 z_2)^{K}}{1-z_1 z_2} & 
\D \frac{(z_2)^{2K}}{1-z_2^2} &  \ldots & \D \frac{(z_2z_N)^K}{1-z_2z_N} \\
	\vdots & \vdots & \vdots & \vdots &  \vdots \\
	\bar{\xi}_N & \D \frac{(z_1 z_N)^K}{1-z_1 z_N} & 
\D \frac{(z_2 z_N)^K}{1-z_2 z_N} & \ldots & \D \frac{z_N^{2K}}{1-z_N^2} \\
	\end{array}\right|,
\eeq
 in terms of the quantities
\beq\label{eq:barxi}
 \bar{I} = I - \sum_{k = 0}^{K-1} g_k^2, \quad  \quad \bar{\xi}_n =\xi_n - \sum_{k=0}^{K-1}g_k z_n^k.
\eeq where:
\bea\label{eq:values} 
g_k&=&\left[\D \frac{1}{k!} \frac{ d^{k}g(z)}{dz^k}\right]_{z=0}, \quad
0\leq k\leq K-1, \nonumber\\
\xi_n&=& g(z_n), \quad  \quad1\leq n \leq N.\eea
In fact,  it can be shown  \cite{Abbas:2010EPJA} that the condition (\ref{eq:gI1}) implies not only  the positivity  (\ref{eq:posit})  of ${\cal D}$, but also the positivity of all its minors.

The inequality (\ref{eq:posit}) defines an allowed domain for the real values $g(z_n)$ of the function at $N$ real points $z_n\in(-1,1)$, and the first $K$ derivatives $g_k$ at $z=0$.  
 In our application we consider $K=2$, noting that the coefficients $g_0$ and $g_1$ entering (\ref{eq:values} ) depend on the charge radius $\langle r^2_\pi \rangle$ defined in Eq. (\ref{eq:taylor}). We further take $N=3$, choosing two points as input\footnote{As discussed in Refs.
 \cite{Ananthanarayan:2012tt, Ananthanarayan:2013dpa}, the inclusion of more input points does not improve automatically the results: indeed, if the input values are known only within  some uncertainties, a saturation is rapidly reached when  more points are included, and the predicted  ranges cannot be further narrowed.}, $t_1=t_s$ and $t_2=t_t$  from the conditions (\ref{eq:val}) and (\ref{eq:mod}),
 while  $t_3$  is an arbitrary point below $\tin$. For $t_1<0$  we have from Eqs.
(\ref{eq:h}) and (\ref{eq:gF11})
\beq\label{xin}
g(z_1)=w(z_1)\, \omega(z_1) \,F(t_1) /\omnes(t_1), \quad z_1=\tilde z(t_1).
\eeq
while for $t_n$, $n=2,3$ we have
\beq\label{eq:gFn}
 g(z_n) = w(z_n)\, \omega(z_n) \,|F(t_n)| /|\omnes(t_n)|,  \quad  z_n=\tilde z(t_n),
\eeq
 where the modulus $|\omnes(t)|$ of the Omn\`es function is obtained from Eq. (\ref{eq:omnes}) by the principal value (PV) Cauchy integral
\beq	\label{eq:modomnes}
 |\omnes(t)| = \exp \left(\frac {t} {\pi} \text{\rm PV} \int^{\infty}_{4 m_\pi^2} dt' 
\D\frac{\delta (t^\prime)} {t^\prime (t^\prime -t)}\right).
\eeq
The condition (\ref{eq:posit}) provides the solution of the extremal problem formulated in the previous section: indeed, it can be written as a quadratic inequality for the unknown modulus $|F(t_3)|$, with coefficients depending on known quantities, from which we obtain upper and lower bounds on the unknown modulus. One can prove that the bounds are optimal and the results remain the same if the $\leq$ sign in the condition (\ref{eq:L2}) is replaced by the equality sign
\cite{Meiman,Duren, Abbas:2010EPJA}.  Moreover,
as shown in Refs. \cite{IC,Abbas:2010EPJA}, for a fixed weight $\rho(t)$ in Eq.(\ref{eq:L2}),  the  bounds depend in a monotonous way 
on the value of the quantity $I$, becoming stronger/weaker when this 
value is decreased/increased. 
The positivity of the minors of ${\cal D}$ provide consistency  constraints on the quantities that enter as input, which ensures that the quadratic equations for the bounds have real solutions. 

%%%%%%%%%%%%%%%%%%%%%%%%%%%%%%%%%%%%%%%%%%%%%%%%%
\section{Input quantities and optimization procedure \label{sec:inputs}}

The first inelastic threshold $\tin$ for the pion form factor is due to the 
opening of the $\omega\pi$ channel which corresponds to $\sqrt{\tin}=m_\omega+m_\pi=0.917\,\gev$. 
We calculate the Omn\`es function  (\ref{eq:omnes}) using as input for $t\le \tin$ the  phase shift $\delta_1^1(t)$ from Refs. \cite{ACGL, Caprini:2011ky} and \cite{GarciaMartin:2011cn}, 
which we denote as Bern and Madrid  phases, respectively.
Above $\tin$  we use  a continuous function $\delta(t)$, 
which approaches asymptotically $\pi$. As shown in Ref. \cite{Abbas:2010EPJA}, if 
this function is Lipschitz continuous, the dependence of the functions $\omnes(t)$ and $\omega(z)$,  
defined in Eqs. (\ref{eq:omnes}) and (\ref{eq:omega}), respectively, on the arbitrary function $\delta(t)$ for $t>\tin$  exactly compensate  each other, leading to 
results fully independent of the unknown phase in the inelastic region.  This
is one of the important strengths of the method applied in this work.

We have calculated the integral defined in Eq. (\ref{eq:L2}) using the BABAR data \cite{BABAR} 
from $\tin$ up to $\sqrt{t}=3\, \gev$,  continued with  a constant value for the modulus 
in the range $3\, \gev \leq \sqrt{t} \leq 20 \gev$,  smoothly continued at higher energies by a $1/t$ decreasing modulus, as predicted by perturbative QCD \cite{Farrar:1979aw,Lepage:1979zb, Melic:1998qr}.
This model is expected to overestimate the true value of the integral: indeed,  
the central BABAR value of the modulus at 3 GeV is equal to 0.066, {\em i.e.,}  while while next-to-leading-order perturbative QCD 
 \cite{Melic:1998qr}  predicts a much lower modulus, equal 
to 0.011 at 3 GeV and 0.00016 at 20 GeV. As emphasized above, 
larger values of $I$ are expected to produce weaker bounds. Therefore, by using
an overestimate of  the high-energy integral, we obtain larger admissible ranges for the resulting modulus, and there is no danger of underestimating the final uncertainties. 
This makes our procedure very robust. 

We have considered weights of the type (\ref{eq:rhogeneric0}). As discussed in Ref. \cite{Ananthanarayan:2012},  weights with a rapid decrease suppress the high-energy part allowing a precise calculation of the integral but lead to a weaker constraint on the class of the admissible form factors. On the other hand, for weights with a slower decrease the condition (\ref{eq:L2}) has a bigger constraining power, but the value of $I$ is more sensitive to the behavior at high energies.    We have adopted finally the weight $\rho(t)=1/t$, for which the contribution of the range above 3 GeV to the integral (\ref{eq:L2}) is only of $1\%$, and the value of $I$ is \cite{Ananthanarayan:2012tt}
\beq\label{eq:Ivalue1}
I=0.578 \pm 0.022, \eeq 
where the uncertainty is  due to the BABAR experimental errors.  In the applications we have used as input for $I$  the central value quoted in Eq.(\ref{eq:Ivalue1}) increased by the  error, which leads to the most conservative bounds due to the monotony property mentioned  above.

As spacelike  input (\ref{eq:val}) we have used one of the most recent experimental determinations\cite{Horn, Huber}
\bea\label{eq:Huber}	
F(-1.60\,\gev^2)= 0.243 \pm  0.012_{-0.008}^{+0.019}, \nonumber \\ 
F(-2.45\, \gev^2)=  0.167 \pm 0.010_{-0.007}^{+0.013}.
\eea
On the timelike axis, it is convenient to take in (\ref{eq:mod}) values $t_t$ from a region of higher energies, where the data are more accurate.  In our study we  used, as in \cite{Ananthanarayan:2013dpa}, input from the region $0.65\gev \leq\sqrt{t_t} \leq 0.70\gev$,  where BABAR  \cite{BABAR, BABAR1} quotes 26 experimental 
points,   KLOE \cite{KLOE3} reports 8, while CMD2 \cite{CMD2:2} and SND \cite{SND}  have each 2 measurements. Input data from higher energies, which can be included in the same way,  are expected to influence the bounds only near the right end of the range considered. 

 Since the Fermi--Watson theorem  (\ref{eq:watson})  is valid only if isospin is conserved, we have worked in the exact isospin limit by correcting the experimental input for the main isospin violating effect in  $e^+e^-\to\pi^+\pi^-$,  due to the $\rho-\omega$ mixing.  More exactly, we have divided the experimental  modulus  by the factor  $|F_{\omega}(t)|$, where
\begin{equation} \label{eq:iso}
F_{\omega}(t)=\Big(1+\epsilon\,\frac{t}{t_\omega-t} \Big), ~~ t_\omega=(m_\omega-i\Gamma_\omega/2)^2, 
\end{equation}
with  $m_\omega=0.7826\,\gev$, $\Gamma_\omega=0.0085\,\gev$ and  $\epsilon=1.9\times 10^{-3}$ \cite{Leutwyler:2002hm, Hanhart:2012wi}. 
After deriving the upper and lower bounds on the form factor in the isospin limit, we have  multiplied them back by the factor  $|F_{\omega}(t)|$ for the calculation of the integral (\ref{eq:amu}). 
 
An important point to bear in mind is that, except for the normalization condition (\ref{eq:taylor}) which is exact, all the input quantities that we use 
are known only with some uncertainty. In fact, the proper treatment of these uncertainties is essential for drawing correct conclusions from our formalism.  Following the discussion given in Ref. \cite{Ananthanarayan:2013dpa}, in our analysis we have 
   varied all the input quantities used for the derivation of the bounds within
their error intervals.  The input quantities are the phase (\ref{eq:watson}), the  spacelike value (\ref{eq:val}), the  timelike value (\ref{eq:mod}) and the  charge radius $\langle r^2_\pi \rangle$. From  the combinations of these values we have generated a large sample of ``pseudoexperimental data", which we have used  as input for the calculation of upper and lower bounds on $|F(t)|$ in the region of interest. Finally, we have taken the most conservative bounds,  {\em i.e.,} the largest upper bound and the smallest lower bound  on $|F(t)|$ from the values obtained with the sample of generated data. 

\begin{figure}[htb]\vspace{0.5cm}
\begin{center}
 \includegraphics[width = 8.cm]{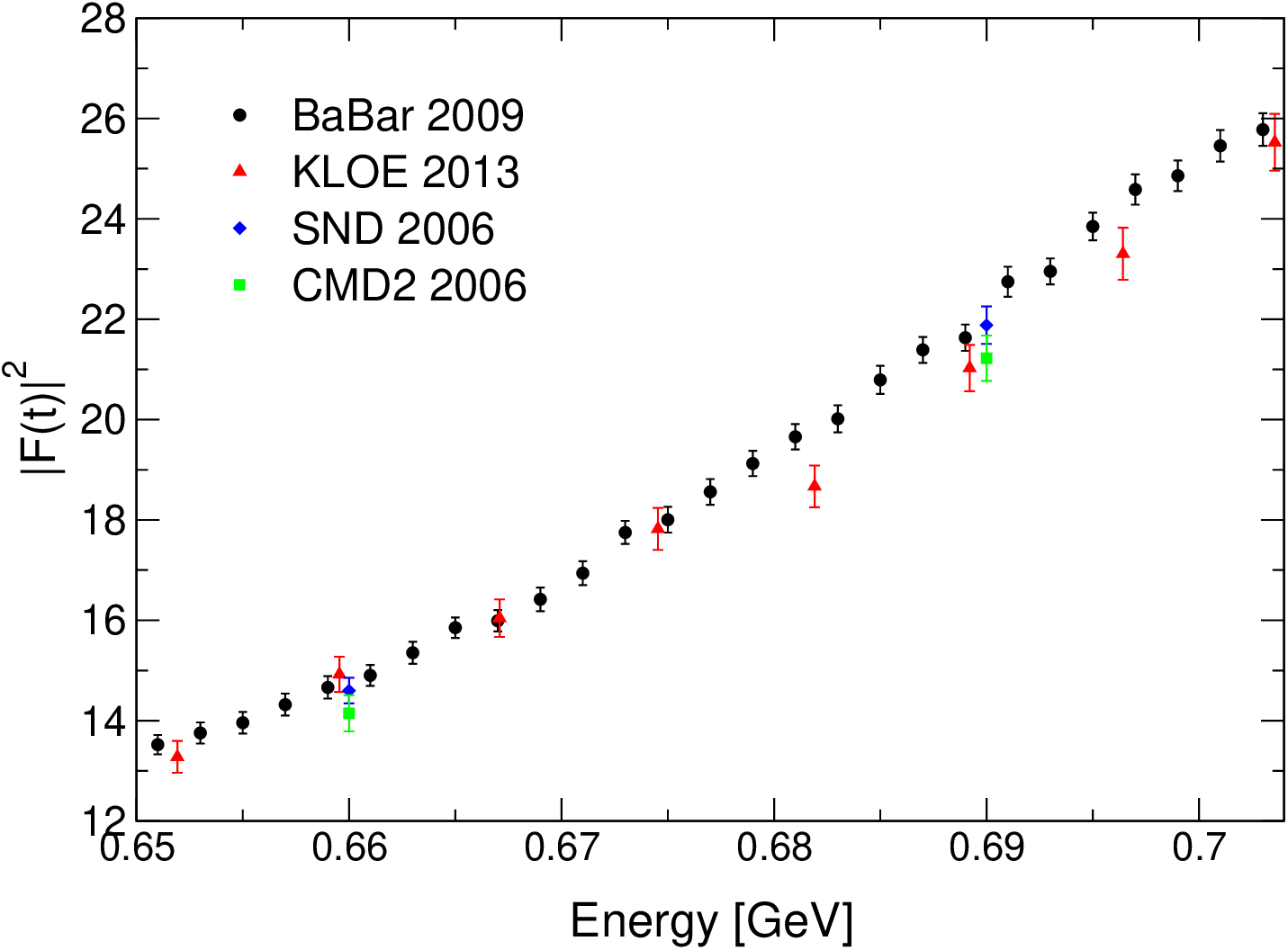}
\caption{Modulus squared $|F(t)|^2$ measured by CMD2, SND,  BABAR and KLOE experiments in the region 0.65-0.70 GeV, used as input in our work. \label{fig:data}}\end{center}
\vspace{0.3cm}
\end{figure}

 The procedure described above has been applied with the input from a fixed spacelike point $t_s$ as given in Eq.(\ref{eq:Huber})  and a fixed timelike point  $t_t$  in  the  range  $0.65\gev \leq\sqrt{t_t} \leq 0.70\gev$ (for completeness the input timelike data are shown in Fig. \ref{fig:data}), with the variation
in the spacelike and the timelike  points taken into account. For each input available at a fixed energy, we have obtained an allowed interval for $|F(t)|$.  Since  the constraints provided by the measurements at different energies must be valid simultaneously, we have taken the ``intersection" of the ranges obtained with fixed spacelike and fixed timelike input, {\em i.e.,}  the smallest upper bound and the largest lower bound  \cite{Ananthanarayan:2013dpa}.  The procedure has been carried out efficiently with a combination of Mathematica and Fortran programs.

Using a similar procedure, we have derived in Ref. \cite{Ananthanarayan:2013dpa}  bounds on the shape parameters at $t=0$, in particular the charge radius $\langle r^2_\pi \rangle$. The intersection of the admissible intervals obtained with input from various energy points led to the range  $(0.42-0.44) \fmsq$ for this quantity.  In the present study, we have allowed the charge radius to vary in the slightly larger range $(0.41-0.45) \fmsq$, which covers practically the allowed intervals obtained with input from individual points. In fact, the positivity of the minors of the determinant (\ref{eq:det}), which is imposed in order to obtain  real bounds on $|F(t)|$ from the inequality (\ref{eq:posit}), amounts to the same restrictions on the radius as  those considered in Ref. \cite{Ananthanarayan:2013dpa}. Therefore,  the input range adopted for $\langle r^2_\pi \rangle$ plays a weak  constraining role on the final bounds on $|F(t)|$.

By applying the procedure described above, we have
computed bounds on $|F(t)|$ for energies between  threshold and  0.63 GeV.  We have repeated the calculation separately with the timelike input from the four $e^+e^-$ experiments  CMD2-2006 \cite{CMD2:2}, SND \cite{SND},  BABAR \cite{BABAR}, and  KLOE-2013 \cite{KLOE3}. In each case we have used as input the two phases \cite{ACGL,Caprini:2011ky} and \cite{GarciaMartin:2011cn}, denoted as Bern and Madrid, respectively.

%%%%%%%%%%%%%%%%%%%%%%%%%%%%%%%%%%%%%%%%%%%%%%%%%%%%%%%%%%%%%%%%%%%%%%%%%%%%%%%%%%%%%%%%%%%%%%%
\section{Results \label{sec:results}}
%%%%%%%%%%%%%%%%%%%%%%%%%%%%%%%%%%%%%%%%%%%%%%%%%%%%%%%%%%%%%%%%%%%%%%%%%%%%%%%%%%%%%%%%%%%%%%%

\begin{figure}[htb]\vspace{0.8cm}
\begin{center}
\includegraphics[width = 8.cm]{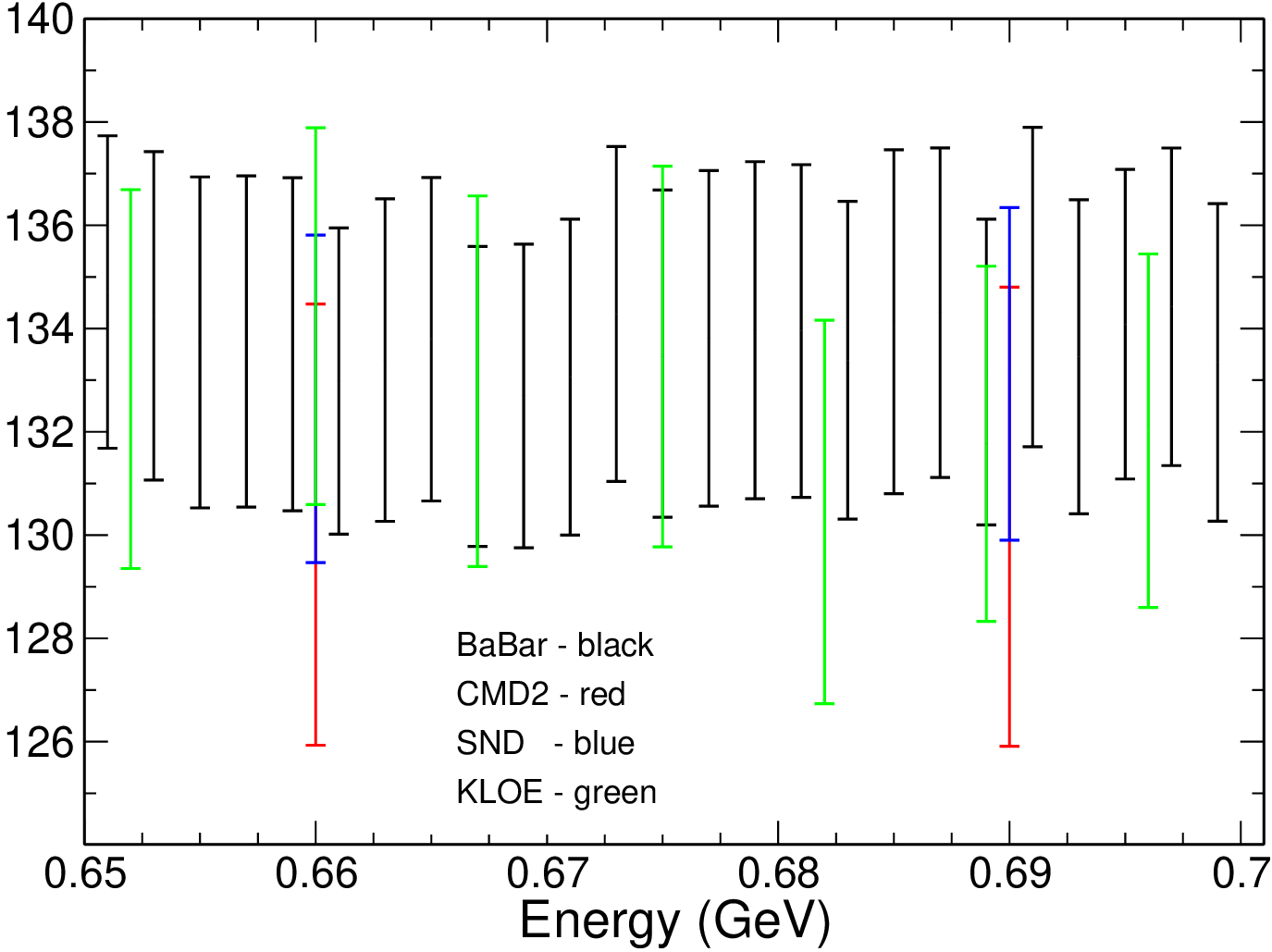}
\caption{Upper and lower bounds on  $a_\mu^{\pi\pi, \LO}\,[0.30 \gev,\, 0.63 \gev] \times 10^{10}$ using as input the 
Bern phase and the timelike modulus measured in the region 0.65-0.70 GeV  by the $e^+e^-$ experiments. \label{fig:all_bern}}\end{center}
\vspace{0.3cm}
\end{figure}

We first illustrate the intersection procedure explained in the previous section considering as an observable the contribution  $a_\mu^{\pi\pi, \LO}\,[0.30 \gev,\, 0.63 \gev]$ to the muon $g-2$. In Fig. \ref{fig:all_bern} we present the allowed ranges for this quantity obtained using  as input one experimental modulus as a function of the energy of the input in the range 0.65 - 0.7 GeV.  More precisely, the intervals are delimited by the upper and lower bounds on the quantity of interest calculated for each input modulus  shown in Fig. \ref{fig:data}, taking into account all the uncertainties  and including the spacelike input as explained in the previous section. The Bern phase from Refs.\cite{ACGL,Caprini:2011ky} was used in this calculation. The final allowed domain is obtained for each experiment as the common part of the corresponding intervals shown in  Fig. \ref{fig:all_bern}.

For the BABAR experiment, we have obtained a large number of such intervals, which are narrower than those obtained with input from the other experiments, due the greater accuracy of the measurements, seen in Fig. \ref{fig:data}. In addition, the intervals are mutually consistent in an impressive way. The allowed intervals obtained with CMD2 and  SND input are larger, reflecting the poorer accuracy of the measurements of these experiments in the range  0.65-0.70 GeV. The intervals obtained with the  KLOE input are slightly larger than those obtained with BABAR and exhibit a more pronounced variation with the input point. As a consequence,  the final range obtained from the common overlap of the individual intervals will be slightly smaller for KLOE than for BABAR. 
This result signals the sensitivity of our machinery to the internal consistency of each data set we have used as our inputs. As our purpose is just to illustrate the power of the formalism,  we have taken the  experimental 
data as such and have kept all the input points as acceptable, 

For the near-threshold integral  $a_\mu^{\pi\pi, \LO}\,[2 m_\pi, \,0.30 \gev]$ the corresponding intervals are almost identical for all experiments and show a great stability with variation of the input, which is explained by the fact that in this region the bounds are less influenced  by the data above 0.65 GeV, and more by the common inputs, {\em i.e.,} the spacelike values, the charge radius and the phase. 

In Tables \ref{table:amu1} and \ref{table:amu2}, we present the results obtained from the intersection of the intervals discussed above.   We write them in terms of a central value and an error, obtained from the average of the upper and lower bounds and half of the interval width, respectively. For completeness we give the results obtained separately with the two phases, Bern and Madrid.

\begin{table}[hbtp]
\centering
\caption{Central values  and errors for the quantity $a_\mu^{\pi\pi, \LO}\,[2 m_\pi,\, 0.30 \gev]\times 10^{10}$  obtained from the bounds on $|F(t)| $ calculated with  input from the four $e^+e^-$ experiments. \label{table:amu1}}\vspace{0.3cm}
\begin{tabular}{l c c}\hline \hline
& Bern phase & Madrid phase \\\hline
CMD2 06 & 0.5528 $\pm$ 0.0089 & 0.5527  $\pm$  0.0092 \\ 
SND 06& 0.5532 $\pm$  0.0083  & 0.5530 $\pm$   0.0086\\
BABAR 09 &0.5534 $\pm$  0.0080 & 0.5533 $\pm$  0.0083\\ 
KLOE 13 & 0.5531 $\pm$ 0.0080 &  0.5530 $\pm$  0.0084\\ \hline\hline
\end{tabular}
\end{table}

\begin{table}[hbtp]
\centering
\caption{Central values  and errors for the quantity   $a_\mu^{\pi\pi, \LO}\,[0.30 \gev,\, 0.63 \gev]\times 10^{10}$ obtained from the bounds on $|F(t)|$ calculated with input from the four $e^+e^-$ experiments. \label{table:amu2}}\vspace{0.3cm}
\begin{tabular}{l c c}\hline \hline
& Bern phase & Madrid phase \\\hline
CMD2 06 & 130.531 $\pm$ 3.955 &  129.739 $\pm$ 4.545  \\ 
SND 06& 132.775 $\pm$ 2.862  & 132.313 $\pm$  2.759 \\
BABAR 09 & 133.732 $\pm$ 1.761 & 133.484 $\pm$ 1.461 \\ 
KLOE 13 & 132.380 $\pm$ 1.721 &  132.086 $\pm$ 1.451 \\ \hline\hline
\end{tabular}
\end{table}

Since the input phases are calculated theoretically by a similar procedure, the two determinations based on them cannot be considered as statistically independent. It is reasonable then to take  as the final prediction, for each experiment, the simple average of the two values given in Tables \ref{table:amu1} and \ref{table:amu2}.  In contrast, the results obtained with input from the four independent experiments, BABAR,  KLOE, SND and  CMD2, are statistically independent and can be combined  with standard techniques for independent   determinations \cite{PDG}. This gives 
\beq\label{eq:average1}
 a_\mu^{\pi\pi, \LO}\,[2 m_\pi,\, 0.30 \gev]=(0.553 \pm 0.004)	\times 10^{-10},\eeq
and
\beq\label{eq:average2}
a_\mu^{\pi\pi, \LO}\,[0.30 \gev,\, 0.63 \gev]=(132.703  \pm 1.018) \times 10^{-10}.
\eeq

To draw a comparison, we note that Ref. \cite{Davier:2009} quotes  for the contribution from the threshold to 0.30 GeV the value $(0.55 \pm 0.01) \times 10^{-10} $  obtained from a  ChPT fit since in this region there are no data. For the contribution from 0.30 to 0.63 GeV, Ref. \cite{Davier:2009} quotes the value $(132.6 \pm 1.3) \times 10^{-10}$, obtained from combined  $e^+e^-\to \pi^+\pi^-$ experiments.

It is of interest to compare in particular the results given by the BABAR experiment, for which good data exist also at low energies.  For the region from 0.30 to 0.63 GeV,  the experiment BABAR measures the contribution to muon $g-2$ as\footnote{We are grateful to Bogdan Malaescu for sending us these values.} $ (133.877  \pm 1.472) \times 10^{-10}$,
the error being determined from $ (0.8605_{\rm stat\,CovMat} \pm 1.1942_{\rm  syst\,CovMat} ) \times 10^{-10}$. 
On the other hand,  from Table \ref{table:amu2} we obtain for BABAR the contribution $(133.608  \pm   1.611  ) \times 10^{-10}$, quite close and with a slightly larger error than the direct determination. This shows  the  remarkable consistency of the BABAR data with the analyticity constraints imposed in this work. We emphasize that these are independent determinations: in the first method  one integrates the data available below 0.63 GeV, while the second method uses data from energies above 0.65 GeV and extrapolates them in a parametrization-free formalism. Their combination gives the best BABAR value:
\beq\label{eq:BABAR}
a_{\mu, \text{BABAR}}^{\pi\pi, \LO}\,[0.30 \gev,\, 0.63 \gev]=(133.755\pm 1.087)\times 10^{-10}. 
\eeq
By combining this value with the determinations with input from CMD2, SND and KLOE given in Table \ref{table:amu2}, we obtain  our final prediction
\beq\label{eq:average3}
a_\mu^{\pi\pi, \LO}\,[0.30 \gev,\, 0.63 \gev]=( 133.083 \pm 0.837) \times 10^{-10}.
\eeq

\begin{figure}[htb]\vspace{0.5cm}
\begin{center}
 \includegraphics[width = 8.cm]{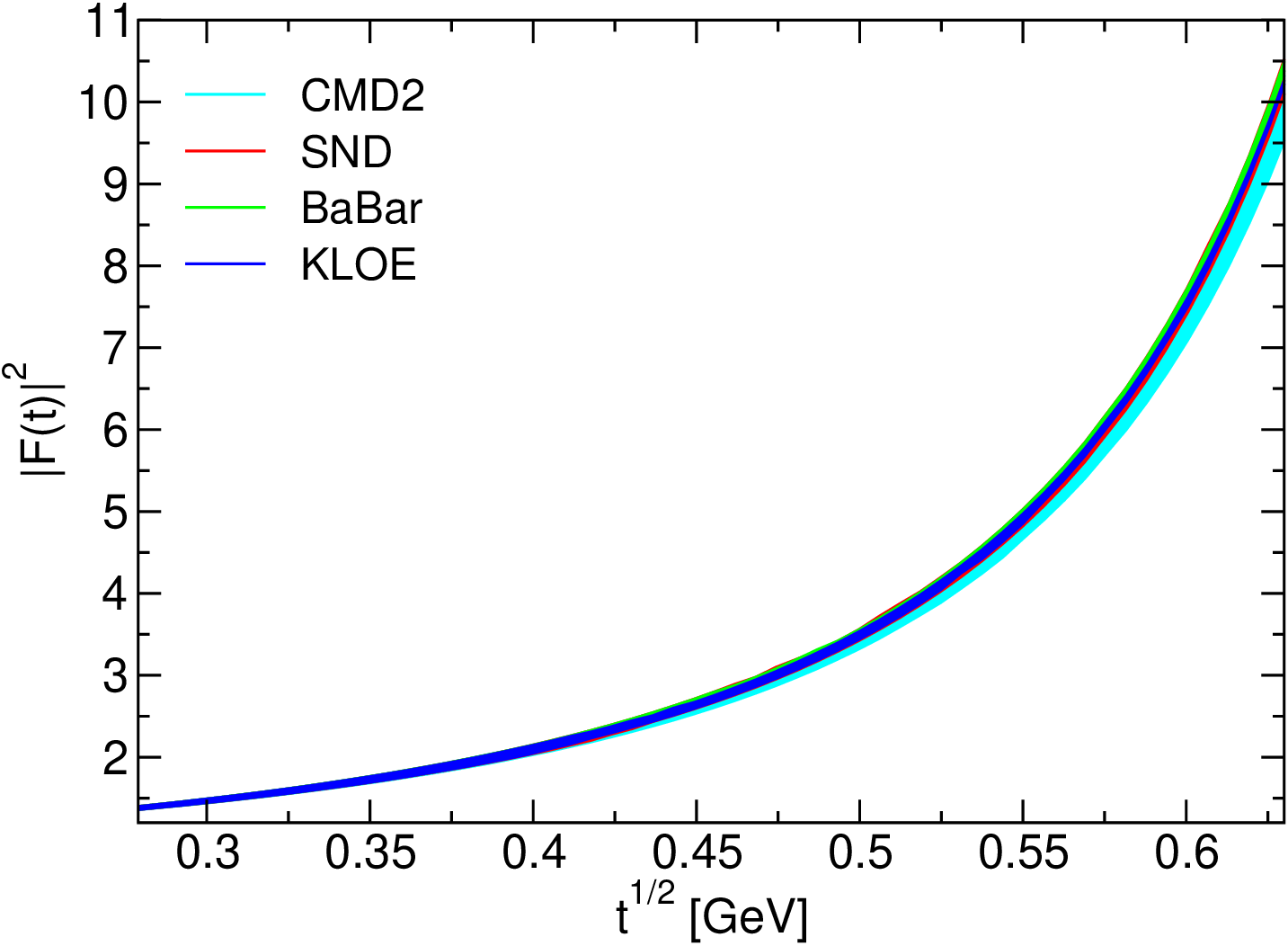}
\caption{Upper and lower bounds on $|F(t)|^2$  in the energy region from threshold to 0.63 GeV,  using  timelike data from the region 0.65-0.70 GeV from CMD2, SND,  BABAR, and KLOE experiments.  \label{fig:epem}}\end{center}
\vspace{0.3cm}
\end{figure}

\begin{figure}[htb]\vspace{0.3cm}
\begin{center}
 \includegraphics[width = 8.cm]{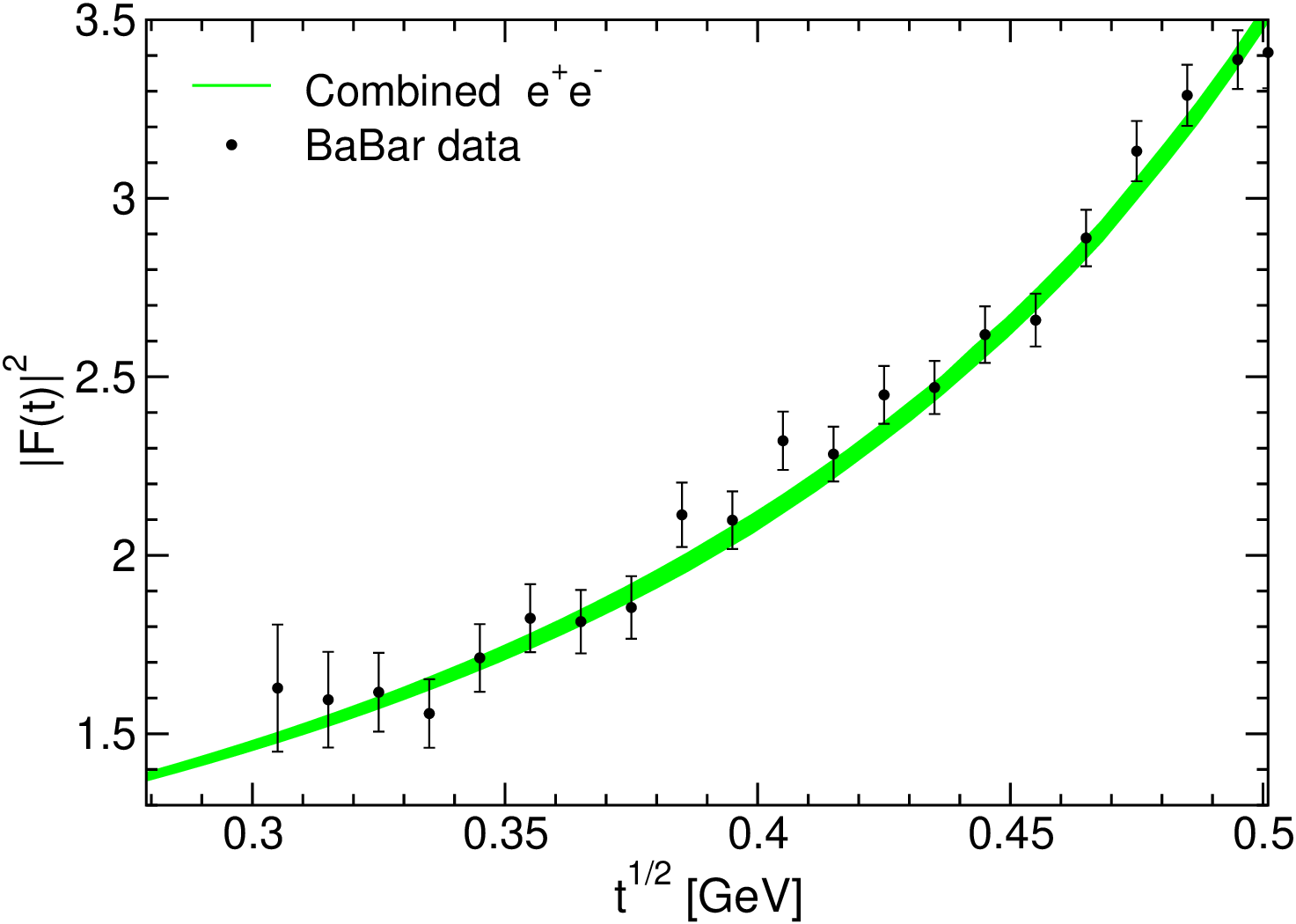}
\caption{Allowed band for $|F(t)|^2$  in the energy region from threshold to 0.5 GeV, obtained by combining  the bounds from 
$e^+e^-$  experiments, compared with the available experimental data.\label{fig:combinedlow}}\end{center}
\vspace{0.3cm}
\end{figure}

\begin{figure}[htb]\vspace{0.3cm}
\begin{center}
 \includegraphics[width = 8.cm]{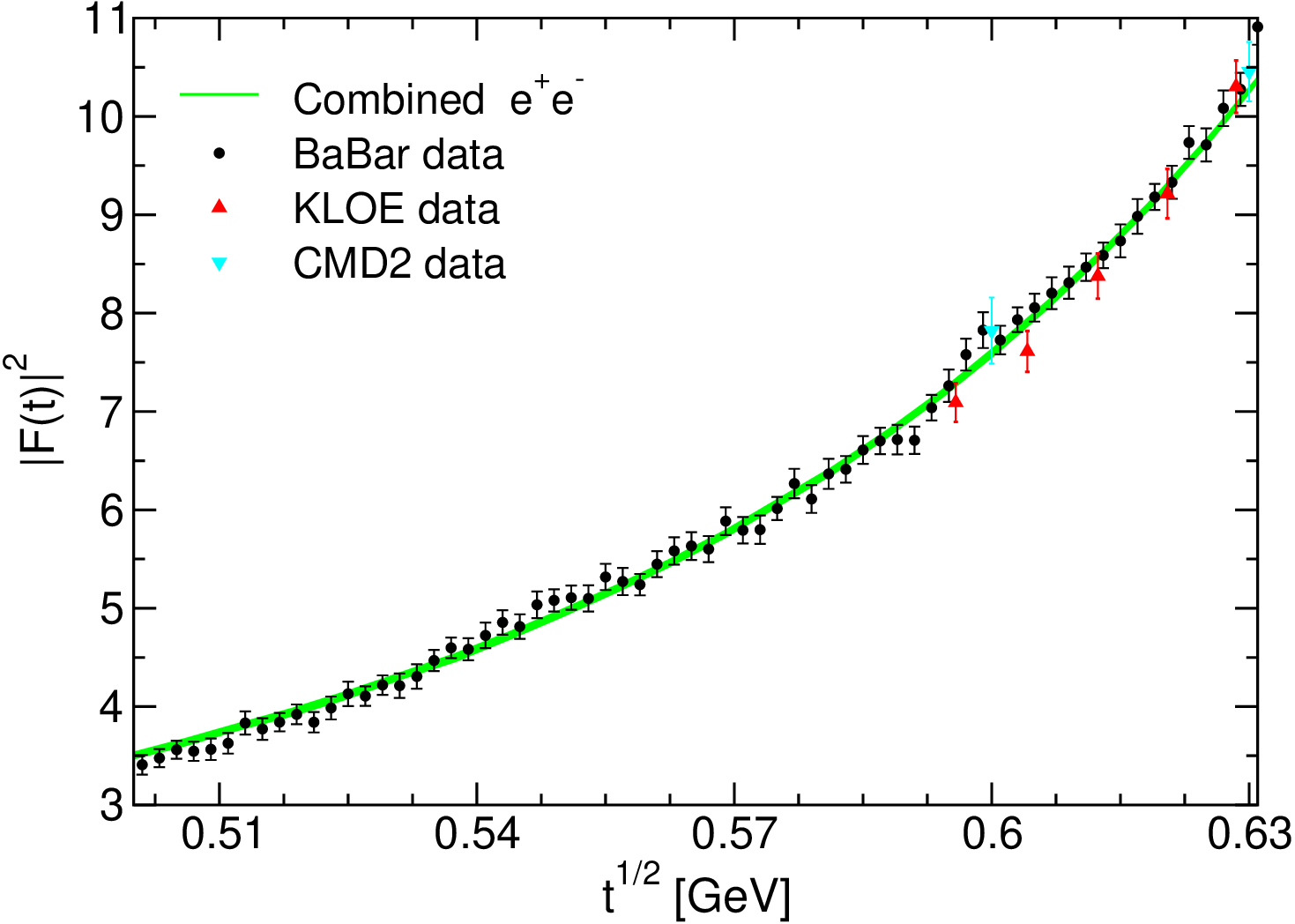}
\caption{Allowed band for $|F(t)|^2$  in the energy region from  0.5 to 0.63 GeV, obtained by combining  the bounds from 
$e^+e^-$  experiments, compared with the available experimental data. \label{fig:combinedhigh}}\end{center}
\vspace{0.3cm}
\end{figure}

It is useful to present also the bounds on the modulus $|F(t)|$ itself at all energies below 0.63 GeV.
In Fig. \ref{fig:epem}  we show the allowed ranges delimited by the upper and lower bounds on  $|F(t)|^2$ in the energy region from threshold to 0.63 GeV obtained with timelike data in the region 0.65-0.70 GeV from  BABAR, KLOE, CMD2 and SND experiments. At each energy the largest interval was obtained using as input  the sample of pseudoexperimental data discussed above; then the intersection of the ranges obtained with the input spacelike and  timelike data was taken. Finally, the average  of the results obtained with Madrid and Bern phases was computed.

The comparison with the similar bounds on the modulus at low energies given in Ref. \cite{Ananthanarayan:2012tt}, which were obtained without the input on the modulus from the elastic part of the unitarity cut, indicates that the additional information on the modulus improves the bounds in  a sizeable way, especially with the data from BABAR and KLOE.  As shown above, the BABAR experiment has many points in this region and  the measurements are consistent  among them so that the intersection of the individual allowed intervals analogous to those shown in Fig. \ref{fig:all_bern} has a little effect in further reducing the domain. For KLOE the relatively narrow domain results mainly from the small common overlap of the individual intervals, as in the previous discussion of $a_\mu$ illustrated in  Fig. \ref{fig:all_bern}. 

By combining the bounds obtained with data from the four $e^+e^-$ experiments, we obtained the final allowed bands for $|F(t)|^2$, shown separately in Figs. \ref{fig:combinedlow} and \ref{fig:combinedhigh}  for the regions from  threshold to 0.5 GeV and from 0.5 to 0.63 GeV.  For comparison we show also the experimental data points available in each region. We note that for KLOE we use the  2013 data published in Ref. \cite{KLOE3}.

From Figs.  \ref{fig:combinedlow}  and \ref{fig:combinedhigh}  it follows that the  allowed ranges for $|F(t)|^2$  calculated here 
are narrower than the error bars, especially at low energies. For the bounds obtained with BABAR and KLOE input, this feature is valid also at higher energies, up to the upper limit of 0.63 GeV.  For completeness we list  in Table \ref{tab:all} the results of the combined bounds on $|F(t)|^2$ for a set of $t$ values, presented in terms of a central value and an error.

\begin{table*}[floatfix]\caption{Central values and errors on $|F(t)|^2$  in the range from threshold to 0.63 GeV, calculated from the  upper and lower bounds obtained with the timelike input from the $e^+e^-$ experiments.\label{tab:all}}
 \begin{center}\begin{tabular}{llll}\hline\hline
  $\hspace{-0.2cm}\sqrt{t}\,(\mbox{GeV})\hspace{0.4cm}$ &  $~~~~~~~|F(t)|^2 $\hspace{3cm} & $\hspace{-0.2cm}
\sqrt{t}\,(\mbox{GeV})\hspace{0.4cm}$ &  $~~~~~~~|F(t)|^2$  \\*[0.1cm]
\hline\\*[-0.4cm]
0.2791 & 1.3803 $\pm $
  0.0096 & 0.4555 & 2.7227 $\pm $
  0.0261 \\
0.2854 & 1.4055 $\pm $
  0.0102 & 0.4618 &  2.8135 $\pm $
  0.0261\\
0.2917&  1.4315 $\pm $
  0.0108 &  0.4681 &  2.9086 $\pm $
  0.0266 \\
0.2980 & 1.4585 $\pm $
  0.0115 &  0.4744 & 3.0168 $\pm $
  0.0270 \\
 0.3043 & 1.4866 $\pm $
  0.0121 &  0.4807 &  3.1258 $\pm $
  0.0274 \\
0.3106 & 1.5159 $\pm $
  0.0128 &  0.4870 &  3.2396 $\pm $
  0.0276\\
0.3169 & 1.5463 $\pm $
  0.0136 &  0.4933 & 3.3628 $\pm $
  0.0260\\
0.3232 &  1.5780 $\pm $
  0.0143 &  0.4996 & 3.4945 $\pm $
  0.0272 \\
0.3295 & 1.6110 $\pm $
  0.0151 &  0.5059 & 3.6363 $\pm $
  0.0288\\
0.3358 & 1.6455 $\pm $
  0.0158 &  0.5122 & 3.7958 $\pm $
  0.0264\\
0.3421 & 1.6813 $\pm $
  0.0167 & 0.5185 &  3.9453 $\pm $
  0.0305\\
0.3484 & 1.7188 $\pm $
  0.0176 & 0.5248 & 4.1173 $\pm $
  0.0313\\
0.3547 & 1.7578 $\pm $
  0.0185 &  0.5311 &  4.3121 $\pm $
  0.0260 \\
0.3610& 1.7988 $\pm $
  0.0195 &  0.5374 & 4.4971 $\pm $
  0.0332\\
0.3673 & 1.8413 $\pm $
  0.0205 & 0.5437  &4.7201 $\pm $
  0.0309\\
0.3736 &1.8860$\pm $
  0.0215 &  0.5500 & 4.9548 $\pm $
  0.0302 \\
0.3799& 1.9329 $\pm $
  0.0226 & 0.5563 &5.2015 $\pm $
  0.0281\\
0.3862 & 1.9819 $\pm $
  0.0236  &0.5626 &5.4681 $\pm $
  0.0289\\
0.3925 &2.0346 $\pm $
  0.0239 &
  0.5689 &5.7562 $\pm $
  0.0292\\
0.3988 &2.0874 $\pm $
  0.0245  &0.5815 & 6.4043 $\pm $
  0.0301 \\
0.4051 &2.1455 $\pm $
  0.0234 &0.5878 &6.7741 $\pm $
  0.0293\\
0.4114 & 2.2059 $\pm $0.0242 &0.5941& 7.1755 $\pm $
  0.0305 \\
0.4177& 2.2692 $\pm $
  0.0232 &0.6004& 7.6231 $\pm $
  0.0340\\
0.4240 &2.3361 $\pm $
  0.0238 &0.6067 & 8.1028 $\pm $
  0.0320\\
0.4303 &2.4058 $\pm $
  0.0248 &0.6130 & 8.6231 $\pm $
  0.0299\\
0.4366 & 2.4775$\pm $
  0.0233 &0.6193 & 9.1781 $\pm $
  0.0293 \\
0.4429 &2.5585 $\pm $
  0.0264  &0.6256 &9.7991 $\pm $
  0.0254\\
0.4492 &2.6365 $\pm $
  0.0253 &0.6319 & \hspace{-0.18cm}10.4738 $\pm $
  0.0176\\
\hline\hline
\end{tabular}
\end{center}\end{table*}

Finally, the values given in Table \ref{tab:all} allow a direct determination of the low-energy contribution to $a_\mu$. We obtain for the near threshold contribution a value identical to that quoted in Eq.(\ref{eq:average1}) and for the contribution from 0.30 to 0.63 GeV the value  $(132.738 \pm 0.949) \times 10^{-10}$, very close to the prediction (\ref{eq:average2}) obtained by combining the values of $a_\mu$, which is a good consistency check of our procedure.

%%%%%%%%%%%%%%%%%%%%%%%%%%%%%%%%%%%%%%%%%%%%%%%%%%%%%%%%%%%%%%%%%%%%%%%%%%%%%%%%%%%%%%%%%%%%%%%%%%%%%%
\section{Discussions and Conclusion \label{sec:conclusion}}

In this paper we have devised an efficient nonperturbative analytic tool  for improving the determination of the pionic contribution to muon $g-2$. The work was motivated by the fact that  the two-pion contribution to the muon magnetic anomaly at low energies is affected by a large relative uncertainty, due to the poor quality of the measured cross sections. Our aim was to improve the accuracy by using additional information on the pion electromagnetic form factor and its  analyticity and unitarity properties. 

The  knowledge of the pion form factor has improved considerably in recent years, from different phenomenological sources: the phase below 0.917 GeV is known through the Fermi--Watson
theorem from the $P$-wave phase shift of $\pi\pi$ scattering, the modulus has been measured 
 by high statistics experiments on $e^+e^-$ annihilation and $\tau$ decays, and  measurements  on the spacelike axis from
electroproduction data  with improved accuracy were also reported.  

We have applied a formalism that exploits in an optimal way the information on the phase below the inelastic threshold, using weak constraints (expressed as an $L^2$-norm condition) on the modulus above the inelastic threshold. 
Combined with the well-known analytic interpolation theory, the formalism is flexible enough to include information at some points inside the analyticity domain  and to lead to upper and lower bounds
on the values at other points. We have used as input the modulus of the form factor measured by the $e^+e^-$ experiments above 0.65 GeV where the accuracy is better, and derived bounds on the modulus below 0.63 GeV, where the experimental data are poor.  A remarkable feature is that, while most of the previous treatments use specific parametrizations to extrapolate to regions not directly accessible to experiment or where the precision of direct measurements is poor, our formalism is parametrization free. As we have mentioned, the price paid for this is that we can derive only bounds on the quantities of interest,  instead of making definite predictions.  Nevertheless, the bounds are very stringent, competing in precision with the present experimental data.  Moreover,  they are independent of the unknown phase above the inelastic threshold. 

Our results for two-pion low-energy contribution  to the anomalous
magnetic moment of the muon, obtained with input from the CMD2, SND, KLOE, and BABAR experiments, are given in  Tables \ref{table:amu1} and \ref{table:amu2}.  The  combined predictions of the $e^+e^-$ experiments are quoted in Eqs.(\ref{eq:average1}) and (\ref{eq:average2}). We have further illustrated with BABAR  how to use precise data at both low and higher energies: by combining the direct integration of the data below 0.63 GeV with our independent determination based on data from higher energies, we obtained the value given in Eq.(\ref{eq:BABAR}), which has a slightly better precision than the separate determinations.

The comparison of the results obtained with input from various $e^+e^-$ experiments given in Table \ref{table:amu2} shows the role played by the accuracy of  the input from higher energies.  As a mathematical exercise,  we investigated the improvement of the bounds after artificially reducing the 
quoted errors of the CMD2 and SND input data. A definite improvement is obtained, but it turns out that it depends  to a large extent  also on the central values of the input, which is explained by the fact that the input quantities  are strongly constrained by analyticity. For KLOE, if the input errors are artificially reduced, the data at various energies quickly become mutually inconsistent and no solution for the bounds is found. 
Our analysis shows that accurate timelike data  on the form factor at intermediate energies are crucial not only for improving the direct contribution to muon $g-2$, but also as ingredients for the analytic extrapolation to lower energies performed with the present method.

Recent work that merits mention is the resonance-based parametrization
of the form factor in a wide range of energies, also including
the same energy ranges of interest to us in this work, using
the same experimental information and wiring in analyticity properties
which is one that is based on a field-theory approach \cite{AK1,AK2,AK3}.
It would be of interest to evaluate
the contribution to the muon $g-2$ from these parametrizations.
It would also be possible to check whether these parametrizations
produce form factors that agree everywhere in the relevant energy
range with the bounds derived in the present work. 

The final result of our analysis, obtained by including  also the BABAR direct measurement, is quoted in Eq. (\ref{eq:average3}).  Compared to the previous determination from combined $e^+e^-$ experiments quoted in \cite{Davier:2009},  it implies  a reduction of the error by $\delta_\mu^{\rm \pi\pi}\sim 5 \times 10^{-11}$. Although not spectacular, the reduction  proves the role of the data from higher energies, exploited by means of analyticity and unitarity,   for improving the determination of the low-energy pionic contribution to muon $g-2$.  Moreover, the confirmation of the central values
is striking as the inputs are completely different.
Our analysis can be considered as provisional: with an improved input from low and intermediate energies, it is expected to further reduce the uncertainty on the hadronic part of the muon anomaly. The application of the method to
data coming from $\tau$ decays is also of interest and will be 
investigated in a separate work.

\subsection*{Acknowledgements} We thank
Bogdan Malaescu for a careful reading of the manuscript and many useful comments
and suggestions. 
I.C. acknowledges support from CNCS-UEFISCDI under Contract  Idei-PCE No 121/2011. D.D. and I.S.I. acknowledge support from Deutsche Forschungsgemeinschaft Research Unit FOR 1873 “Quark Flavour Physics and Effective
Theories”.

\end{document}